\documentclass[10pt,notitlepage,aps,prl,reprint,superscriptaddress,amsmath,amssymb,floatfix,showpacs]{revtex4-1}

\usepackage{float}
\usepackage[utf8]{inputenc}

\usepackage{graphicx}
\usepackage{dcolumn}
\usepackage{bm}
\usepackage[english]{babel}
\usepackage{textcomp}
\usepackage[font=scriptsize]{caption}
\usepackage{subcaption}
\usepackage{amsmath}
\begin{document}
\preprint{APS/123-QED}

\title{Unconventional superconductivity in the nickel-chalcogenide superconductor, TlNi$_2$Se$_2$}

\author{E. Jellyman}
\email{EXJ001@bham.ac.uk}
\affiliation{School of Physics and Astronomy, University of Birmingham, Birmingham B15 2TT, United Kingdom}
\author{P. Jefferies}
\affiliation{School of Physics and Astronomy, University of Birmingham, Birmingham B15 2TT, United Kingdom}
\author{S. Pollard}
\affiliation{School of Physics and Astronomy, University of Birmingham, Birmingham B15 2TT, United Kingdom}
\author{E. M. Forgan}
\affiliation{School of Physics and Astronomy, University of Birmingham, Birmingham B15 2TT, United Kingdom}
\author{E. Blackburn}
\affiliation{School of Physics and Astronomy, University of Birmingham, Birmingham B15 2TT, United Kingdom}
\altaffiliation{Division of Synchrotron Radiation Research, Lund University, SE-22100 Lund, Sweden}
\author{E. Campillo}
\affiliation{Division of Synchrotron Radiation Research, Lund University, SE-22100 Lund, Sweden}
\author{A. T. Holmes}
\affiliation{European Spallation Source ERIC, Box 176, SE-22100 Lund, Sweden}
\author{R. Cubitt}
\affiliation{Institut Laue Langevin, 71 Avenue des Martyrs, 38000 Grenoble, France}
\author{J. Gavilano}
\affiliation{Paul Scherrer Institute, 5232 Villigen, Switzerland}
\author{Hangdong Wang}
\affiliation{Department of Physics, Zhejiang University, Hangzhou 310027, China}
\affiliation{Department of Physics, Hangzhou Normal University, Hangzhou 310036, China}
\author{Jianhua Du}
\affiliation{Department of Physics, Zhejiang University, Hangzhou 310027, China}
\author{Minghu Fang}
\affiliation{Department of Physics, Zhejiang University, Hangzhou 310027, China}
\affiliation{Collaborative Innovation Centre of Advanced Microstructure, Nanjing 210093, China}


\date{\today}

\begin{abstract}
 
We present the results of a study of the vortex lattice (VL) of the nickel chalcogenide superconductor TlNi$_2$Se$_2$, using small angle neutron scattering.  This superconductor has the same crystal symmetry as the iron arsenide materials.  Previous work points to it being a two-gap superconductor, with an unknown pairing mechanism.  No structural transitions in the vortex lattice are seen in the phase diagram, arguing against $d$-wave gap symmetry.  Empirical fits of the temperature-dependence of the form factor and penetration depth rule out a simple \emph{s}-wave model, supporting the presence of nodes in the gap function. The variation of the VL opening angle with field is consistent with earlier reports of of multiple gaps.
\end{abstract}


\maketitle

\twocolumngrid
\section{Introduction}

Nickel-chalcogenides are a new class of superconductor \cite{Wang13,Hong14,Wang15,Goh14,Xu15,Neilson12}, with TlNi$_2$Se$_2$  synthesised in single crystal form and characterised in 2013 by Wang \emph{et al.} \cite{Wang13}. TlNi$_2$Se$_2$ becomes superconducting below $T_c = 3.7$ K, and has been identified as a moderately heavy fermion material with an effective mass of $m^* = (14-20) m_e$~\cite{Wang13}. However, ARPES data~\cite{Xu15} suggests that it is not a strongly correlated material and that the large density of states at the Fermi level results from a van Hove singularity, arising from a quirk in the band structure. There is conflicting evidence as to the nature of the pairing mechanisms in this material \cite{Wang13, Hong14}. Thermal conductivity data and deviations from the Wiedemann-Franz law \cite{Graf96, Durst00} do not support a \emph{d}-wave interpretation. However, the heat capacity in the mixed state shows a power-law dependence of the Sommerfeld coefficient: $\gamma_N \propto B^{0.5}$ \cite{Wang13}. This is typically associated with \emph{d}-wave superconductors \cite{Wright99, Yang01, Meulen90}. In the normal state, TlNi$_2$Se$_2$ shows Pauli paramagnetism \cite{Wang13}. Additional evidence from the heat capacity and thermal conductivity \cite{Hong14} points to a two-gap model with the lower gap suppressed above $B^* \simeq 0.36B_{c2} = 0.29$ T.  The two gaps are estimated to be $\Delta_1 = 0.84k_B T_c$ and $\Delta_2 = 2.01k_B T_c$~\cite{Wang13, Hong14}. To investigate this further, we have undertaken a survey of the vortex lattice (VL) using small-angle neutron scattering (SANS), which can give information about the temperature and field-dependence of the superconductivity in this material.

\begin{figure}
    \includegraphics[width=0.3\textwidth]{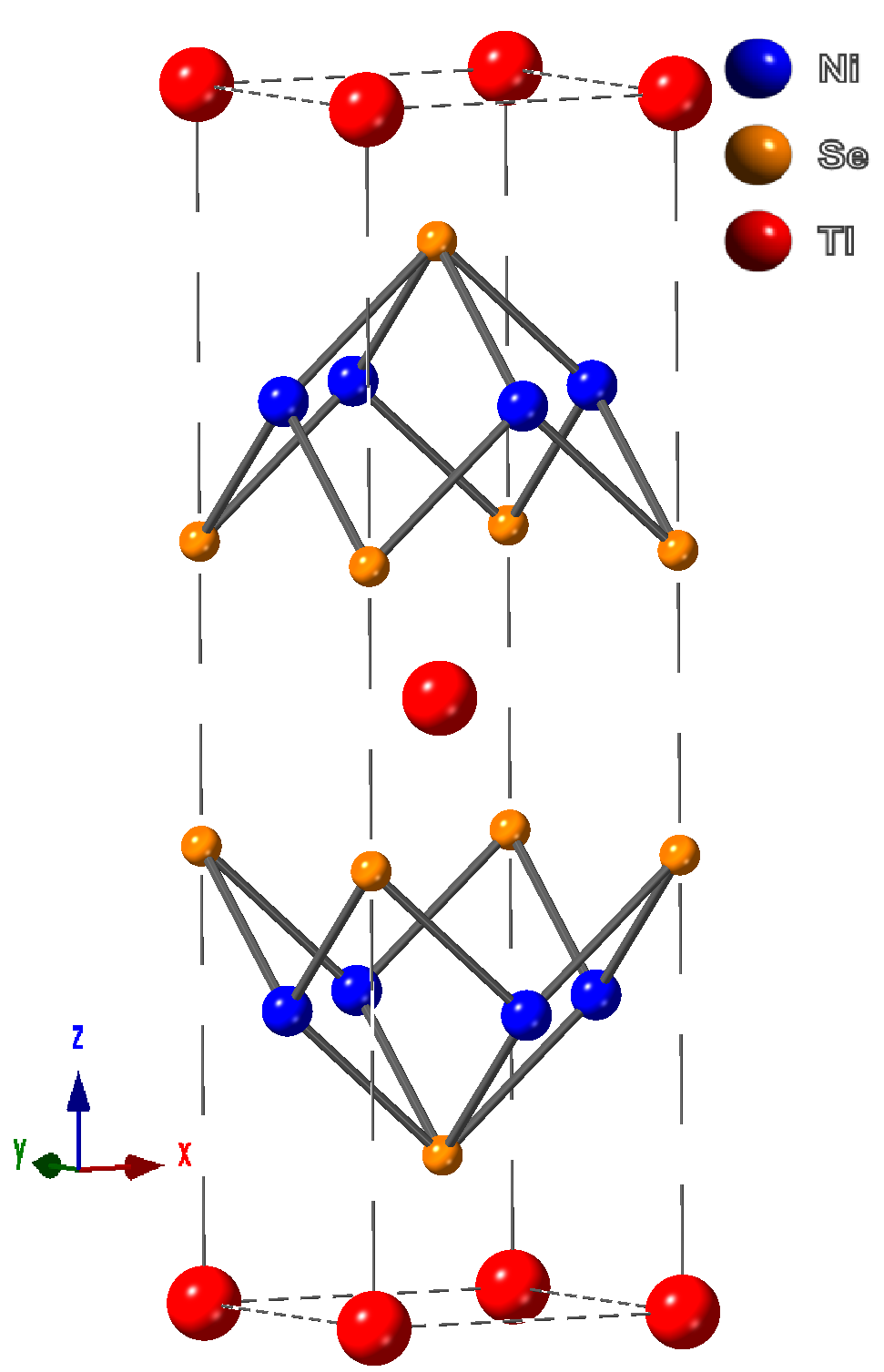}
  \caption{\label{fig:TlNi2Se2Structure} The crystal structure of stoichiometric TlNi$_2$Se$_2$ \cite{Wang13}.}
\end{figure}

TlNi$_2$Se$_2$ has a tetragonal structure (Figure~\ref{fig:TlNi2Se2Structure}), with lattice parameters $a = 3.870\pm 0.001~$\AA${ }$ and $c = 13.435\pm 0.001~$\AA${ }$. It belongs to the \emph{I4/mmm} space group, like the iron-arsenides and CeCu$_2$Si$_2$~\cite{Takenaka17, Steglich79} (the first heavy-fermion superconductor discovered). The resistivity has an anisotropy ratio
of $\rho_c/\rho_{ab} = 1.57$ \cite{Wang13}; this implies that the ratio of effective masses in the {\bf c} and basal directions  $\Gamma_{ac}$ is also $\sim 1.57$. In \cite{Wang13} the ratio of electron mean free path to coherence length is estimated to be $l_e/\xi_0 = 33.3 \gg 1$ using $\xi_0 = 20.3$ nm and $l_e = 677$ nm. We obtain a slightly larger coherence length, but confirm that samples are in the clean limit \cite{Tinkham96}.

As this material is structurally equivalent to the heavy-fermion CeCu$_2$Si$_2$ and the highly anisotropic, multiband KFe$_2$As$_2$~\cite{Kawano11, Kuhn16}, further investigation into TlNi$_2$Se$_2$ could highlight shared characteristics of the \emph{I4/mmm}, 122 chemical structure superconductors.

\section{Experimental Details}

The work presented here was performed  on the D33 instrument at the Institut Laue-Langevin (ILL) \cite{Jellyman16}. Preliminary studies were carried out at SANS-I at the Paul Scherrer Institute (PSI). 

The neutron wavelength used was 7 \AA${ }$ for $B > 0.15$ T and 12 \AA${ }$ for $B \leq 0.15$ T with a bandwidth of $\Delta\lambda/\lambda = 0.1$. The collimation was set to 12.8 m, with a 2D multidetector 12 m from the sample. The sample was mounted in a 17 T horizontal-field cryomagnet equipped with a dilution insert \cite{Holmes12} and was illuminated with neutrons through an aperture of area $ 1.08\times 10^{-4}$ m$^2$. The angle between the sample {\bf c}-axis and the field direction could be altered \emph{in situ} by rotation of the sample by an angle $\Omega$ about the vertical axis.

\begin{figure}[t!]
\begin{subfigure}[t]{0.2\textwidth}
   \includegraphics[width=\textwidth]{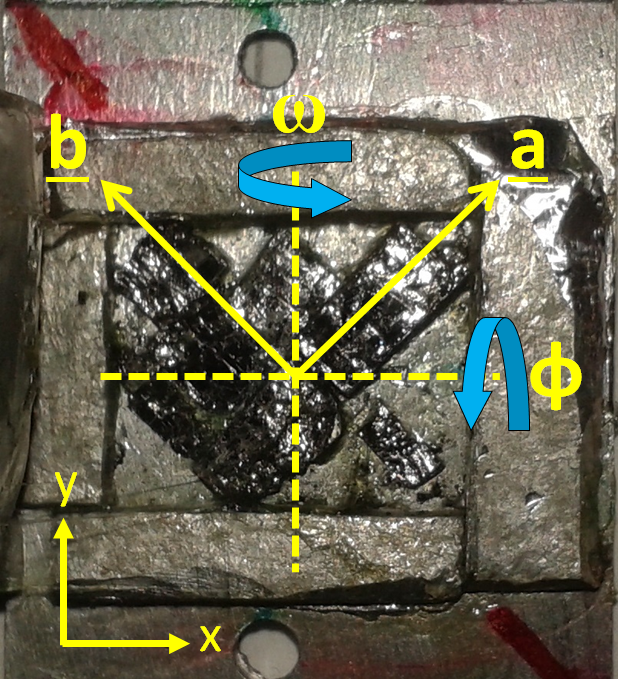}
  \caption{}
	\label{fig:TNS_Mosaic}
\end{subfigure}
\begin{subfigure}[t]{0.25\textwidth}
     \includegraphics[width=\textwidth]{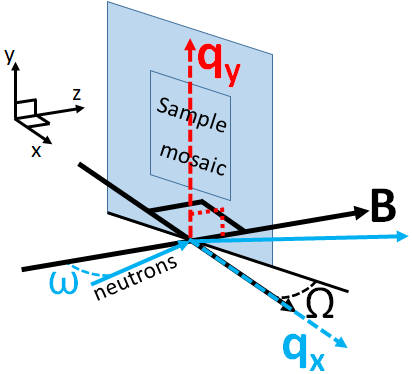}
  \caption{}
	\label{fig:TNS_angles}
\end{subfigure}
\caption{
(a) Image of the mosaic of seven single crystals making up the sample.  They are $\sim 0.13$~mm thick and have a total volume of 4.68~(mm)$^3$.  The solid lines indicate the $ab$-plane alignment of the crystals, at $45^{\circ}$ to the $xy$-axes.  The $c$ axes point out-of-plane.  The dashed lines indicate the axes about which magnet and sample may be rotated together to produce ``rocking curves" of diffracted intensity versus angle as the vortex lattice diffraction spots move through the Bragg condition. The $\omega$ rotation about a vertical axis is illustrated in (b); the $\phi$ rotation is similar but is about the horizontal axis perpendicular to \textbf{B}.
(b) Orientation of the sample plate with respect to the field ($\textbf{B}$) and neutron beam. An $\omega$ rotation is shown, but exaggerated in magnitude for clarity. $\Omega$ is the fixed rotation of the sample \textbf{c}-axis with respect to $\textbf{B}$.The scattering vectors of the diffraction spots will lie in the $q_x-q_y$ plane perpendicular to the field. }
\label{fig:setup}
\end{figure}

A mosaic of seven single crystals was prepared (Figure~\ref{fig:TNS_Mosaic}). The $\bf{c}$ axes were initially parallel to the field, $\textbf{B}$. The magnet and sample inside it could be rotated as a whole to give the small angles between field and neutron beam required to bring the vortex lattice into the Bragg condition for diffraction. The (symmetrically) equivalent  $\textbf{a}$ and $\textbf{b}$ axes were aligned at 45$^{\circ}$ to the vertical axis. When the sample was realigned relative to $\textbf{B}$ by rotation about the vertical axis, the symmetry between the horizontal and vertical crystal directions was broken. This permitted a single vortex lattice (VL) domain to be selected, while two domains were visible with \textbf{B} parallel to \textbf{c}.    

To prepare the vortex lattice at a given temperature and field, the sample was cooled in an applied field through $T_c$ to the target temperature. During cooling, the magnitude of the field was oscillated by $\pm$5~mT about its final value. This procedure is known to improve the structural perfection of the VL \cite{White09}, particularly at low fields. For temperature scans, data were collected by raising the temperature from base, rather than warming and cooling through $T_c$ for each point. 

\begin{figure}[t!]
\begin{subfigure}[t]{0.2275\textwidth}
    \includegraphics[width=\textwidth]{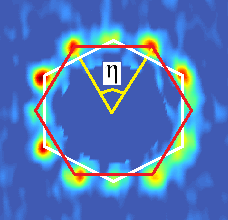}
  \caption{}
	\label{fig:0p3T_0p1K}
\end{subfigure}
\begin{subfigure}[t]{0.22\textwidth}
    \includegraphics[width=\textwidth]{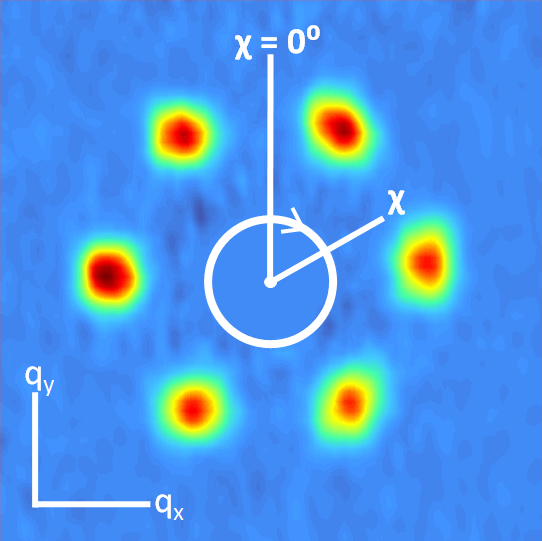}
  \caption{}
	\label{fig:0p4T_0p1K}
\end{subfigure}
\caption{Diffraction patterns of the VL obtained from $\omega$ and $\phi$ rocks of $\pm$0.8$^\circ$ in steps of 0.05$^\circ$ at 2 minutes per point at $B$ = 0.25 T and $T$= 130 mK. These diffraction patterns represent the sum of the measurements of intensity due to the VL as the cryostat, field and sample are rotated through $\phi$ and $\omega$. In panel (a), the $c$ axis is parallel to $\textbf{B}$ ($\Omega = 0^{\circ}$).  Two vortex lattice domains are visible, illustrated by the red and white hexagons. The opening angle $\eta$ is used to study the lattice anisotropy.  In panel (b), $\Omega = 30^{\circ}$. One domain is now dominant (the one marked by the red hexagon in panel (a)).  In panel (b) we show the reciprocal $q_x$ and $q_y$ coordinates, alongside the azimuthal angle $\chi$. This angle is used to denote the angular position of a VL spot in the diffraction image as well as describe relative shifts in the position of VL spots with respect to each other. Image (b) has undergone a Bayesian statistical treatment \cite{Holmes14} to improve the signal to noise ratio of the diffraction pattern.}
\label{fig:san_scan}
\end{figure}


At set values of field ($B$) and temperature ($T$), the diffraction pattern was collected by rocking through the angles $\omega$ and $\phi$ as described in Fig.~2. Background scans were taken in the normal state at $T>T_c$  and subtracted from the VL foreground measurements. The resulting diffraction patterns were analysed using the software package GRASP~\cite{Dewhurst03}. Figure \ref{fig:san_scan} gives examples of such diffraction patterns.

The VL was measured with $\textbf{c}$ at angle $\Omega$ =  $0^{\circ}$, $10^{\circ}$ and $30^{\circ}$ to $\textbf{B}$, where the nonzero angles select one VL domain. Field dependent measurements were taken over the range 0.05 T to 0.5 T at 130 mK. Temperature dependent measurements were taken over the range of 90 mK to 1.85 K. Backgrounds for both temperature- and field-dependence were taken at 4 K at 0 T, 0.15 T and 0.5 T using the same neutron wavelength as for the foreground data at each field.

\begin{figure}
    \centering
    \includegraphics[width=0.48\textwidth]{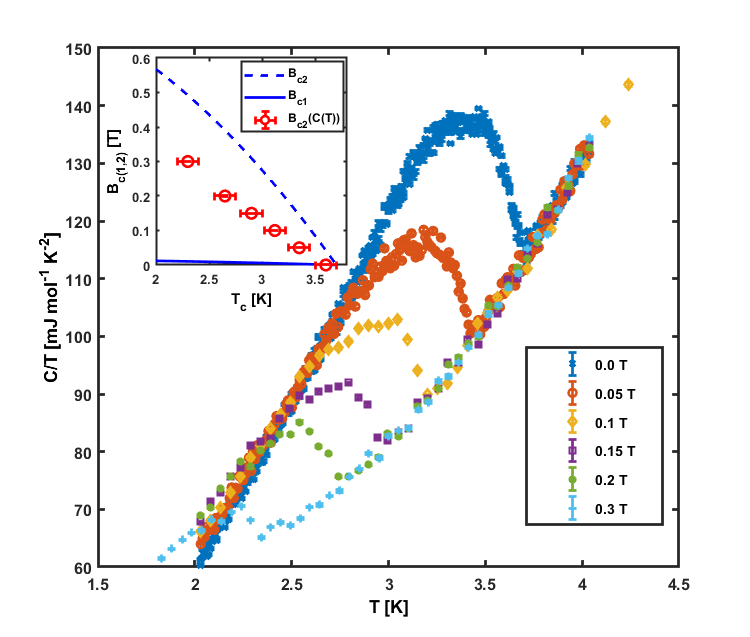}
    \caption{\label{fig:CT_with_B_c_2} Heat capacity divided by temperature versus temperature for various values of field that were used during the neutron scattering investigation. The inset shows the extracted $B_{c2}$ and $T_c$ values from our $C(T)$ investigation alongside the Ginzburg-Landau curves for $B_{c_1}$ (solid line) and $B_{c2}$ (dashed line) based on the values in \cite{Wang13, Hong14}.}
\end{figure}

Additionally, heat capacity measurements were conducted at the University of Birmingham to confirm the relationship between $B_{c2}$ and $T$ for the samples shown in Fig.~2. This investigation was performed on a Physical Properties Measurement System (PPMS) in the ranges $1.8$ K $ < T < 4.25$ K and $0$ T $<B<0.3$ T. The results are presented in Figure~\ref{fig:CT_with_B_c_2}. The inset shows that our measured critical fields and temperatures are lower than the values given by \cite{Wang13, Hong14}. This discrepancy arises because Refs.~\onlinecite{Wang13} and \onlinecite{Hong14} use the \emph{onset} rather than midpoint of the heat capacity transition, and there is actually little difference between the samples. In what follows, we use our values as more representative of the bulk. Our heat capacity results were limited by the 1.8 K lower limit of the PPMS, but with a fit we obtain the following parameters from them: $B_{c2}(0$ K$) = 0.48 \pm 0.03$ T, $T_c(0$ T$) = 3.55 \pm 0.05$ K. From this we generate our estimate of the coherence length: $\xi_0 = (26.1 \pm 1.6)$ nm.  For our temperature-dependence measurements at $\Omega = 30^{\circ}$.  We estimate $B_{c2}(0$ K$, \Omega = 30^{\circ}) = 0.52 \pm 0.03$ T and $\xi_0(\Omega = 30^{\circ}) = 25.1 \pm 1.5$ nm.  We use these values of $B_{c2}$ and $\xi_0$ for the analysis of the temperature dependence of $\langle|F(q,T)|\rangle$ and $\lambda(T)$. In these measurements at an applied field of 0.15 T the transition to the normal state is at $T = 2.8\pm0.1$ K (midpoint of the heat-capacity jump). This was used as ``$T_c$" in analyzing the $T$-dependent neutron scattering investigations at this field.


\section{Results}

\subsection{Vortex lattice structure}


Unconventional pairing mechanisms, multi-gap situations and heavy fermion behaviour are often accompanied by some form of VL structural change, such as the transition from hexagonal to square/rhombic VL commonly seen in \emph{d}-wave systems ~\cite{Kawano11, Kawano13, Kuhn16, Bianchi08}, which can be observed directly from the diffraction patterns as a function of temperature or field. 

The presence of multiple band gaps in a superconductor can be demonstrated by the field-dependence of superconducting properties, such as the anisotropy~\cite{Kuhn16}. There are two intrinsic sources of anisotropy in a superconductor; angular variations in the Fermi velocity, ${\bf v}_{F}$, over the Fermi surface sheets that carry the Cooper pairs, and/or in the energy gap, $\Delta$.

Considering first the Fermi velocity, if the field is applied parallel to the {\bf c}-axis, the VL is sensitive to anisotropies in the $a-b$ plane.  TlNi$_2$Se$_2$ has tetragonal symmetry, and so we expect isotropic behaviour, unless ``nonlocal'' effects are significant~\cite{Kogan97}.  These can give rise to preferred VL orientations, as well as distortions away from the perfect hexagonal lattice as the field is increased.

When the field is rotated relative to the {\bf c}-axis this introduces anisotropy as the {\bf a}- and {\bf c}-directions are inequivalent.  This can give rise to a distorted VL structure, as in KFe$_2$As$_2$ \cite{Kuhn16}.  For TlNi$_2$Se$_2$, unlike the isotructural KFe$_2$As$_2$, the electronic structure is fairly 3-dimensional, with an effective mass ratio $\Gamma_{ac} = [\langle v_F^2(a)\rangle/\langle v_F^2(c)\rangle] \sim 1.57$~\cite{Wang13}, so this effect will be less strong.

The other source of anisotropy is the superconducting gap $\Delta$. Just like the Fermi velocity anisotropy, gap anisotropy can cause VL structure distortions and phase transitions \cite{Suzuki10,Affleck97,Franz97}.  The most obvious effect would be from the presence of nodes.  

In a multi-band superconductor, both $\Delta$ and ${\bf v}_{F}$ may vary within a single sheet, but their behavior is likely to be different on different bands.  The experimentally observed anisotropy will have a value intermediate between those of the separate Fermi sheets~\cite{Kuhn16, Cubitt03}.  However, the application of field may affect these sheets differently (for example, by closing the gap on one sheet), giving rise to a field-dependent anisotropy.

Anisotropy in the VL gives rise to departures of the opening angle (see Figure~\ref{fig:0p3T_0p1K}) from the isotropic value of $\eta = 60^{\circ}$, and concomitantly, in the lengths of the wavevectors, which also depend on the value of the field.  Here, we have collected data from all of the first order Bragg spots, and focus on the opening angle, which we find to be the more accurate measure of the anisotropy than the wavevector magnitudes.  

\begin{figure}
\begin{center}
    \includegraphics[width=0.45\textwidth]{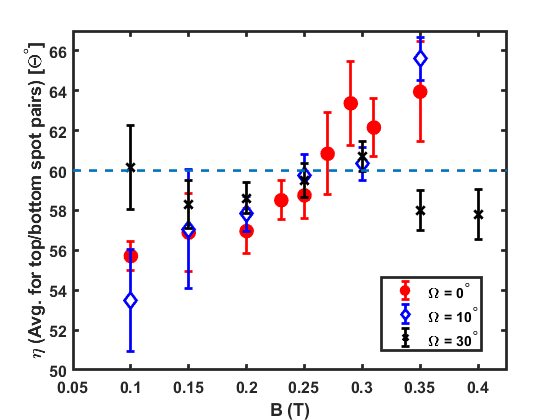}
  \caption{\label{fig:ETA_average} The average opening angle $\eta$ for the top and bottom spots versus applied field. The data were taken at a temperature of 130 mK for three different angles $\Omega$ between the applied field and the crystal {\bf c}-axis. A deviation from 60$^{\circ}$ indicates a contraction or expansion of the hexagonal VL along the horizontal axis and the opposite effect along the vertical axis.}
\end{center}
\end{figure}

Figure~\ref{fig:ETA_average} illustrates the average angular separation as a function of applied field between the pairs of top and bottom spots in the VL diffraction pattern, specifically in the domain marked by the red hexagon in Figure~\ref{fig:0p3T_0p1K}, for different values of $\Omega$. 
Our results show a weak linear relationship with $B$, passing through the isotropic value of $\eta = 60^{\circ}$, with the slope of this response changing dramatically at large $\Omega$.  At $\Omega = 0^{\circ}$, the field variation in $\eta$ must arise primarily from nonlocal effects, but suppression of a gap may also play a role.  On rotating to $\Omega = 30^{\circ}$, the small $ac$ anisotropy introduced should affect the VL distortion; this is expected to increase $\eta$ by $\sim 3^{\circ}$.  It therefore appears that the apparent isotropy at this $\Omega$ is an accidental cancellation of all of the effects described above.

Refs.~\cite{Wang13} and \cite{Hong14} identify the material as a nodeless, two-gap  heavy fermion superconductor with potentially unconventional pairing mechanisms. We observe no significant structural transitions or discontinuities in the VL signal, as a function of either $\Omega$ or $B$, other than in Figure~\ref{fig:ETA_average} at $\approx 0.26$ T; here the opening angle crosses over from more acute to more obtuse than $60^{\circ}$. The specific absence of VL structural transitions is suggestive, but does not completely rule out the unconventional case for this material. 

\subsection{Integrated intensity and form factor}

The form factor, $F(q)$ is a measure of the spatial variation of the field inside the VL relative to the average field.  The latter will be very close to the applied field in our case, due to our plate-like geometry, and because our applied fields are much larger than the lower critical field. The form factor may be calculated from the Christen formula~\cite{Christen77} which relates it to the integrated intensity,
\begin{equation}
I(q) = 2\pi V \lambda_n^2 \phi_n \left(\frac{\gamma_n}{4}\right)^2 \frac{|F(q)|^2}{\Phi_0^2 \cos(\zeta) q},
\label{eq:Christen}
\end{equation}
where $V$ 
 is the volume of the sample mosaic occupied by the VL domain being measured, $\lambda_n$ is the neutron wavelength, $\phi_n$ is the neutron flux (extracted via a direct beam measurement with known aperture size ($\phi_n= 7.71\times10^9$ m$^{-2}$s$^{-1}$ at $\lambda_n$ = 7 \AA~ and $\phi_n= 7.81\times10^8$ m$^{-2}$s$^{-1}$ at 12 \AA),
  $\gamma_n = 1.92$ is the gyromagnetic ratio of the neutron,
  $\Phi_0$ is the flux quantum, $q$ is the magnitude of the momentum transfer for the relevant spots in the diffraction pattern and $\zeta$ is the Lorentz angle (the angle between the {\bf q} of the spot being analyzed and the normal to the rocking angle axis).

The integrated intensity for a given VL Bragg peak is obtained by measuring the scattered neutrons as a function of rocking angle ($\omega$, $\phi$) through the Bragg peak.  After subtraction of an averaged background, 
the resulting rocking curve is fitted using a Lorentzian function and $I(q)$ is the area under the fitted curve.  All of the individual integrated intensities from the 6 spots of a given VL domain are then averaged to give $\langle I(q) \rangle$ for that domain; this is then used to calculate the form factor using Eq.~\ref{eq:Christen}.

\begin{figure}
    \includegraphics[width=0.48\textwidth]{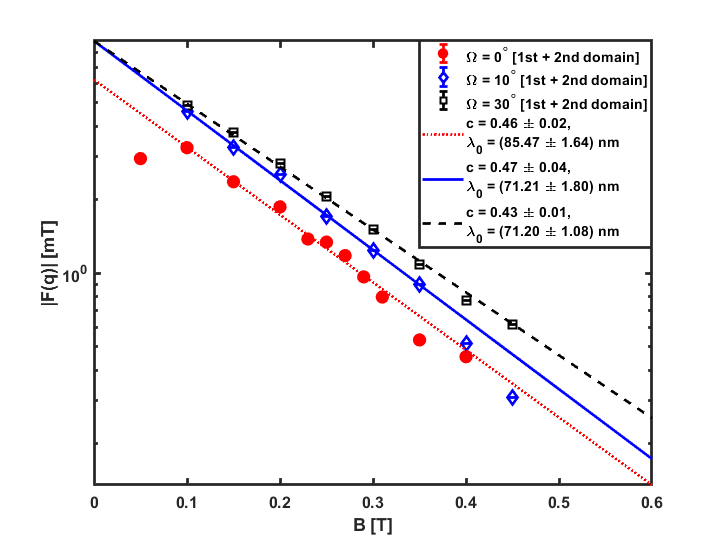}
  \caption{\label{fig:FF_B_pub_log}Form factor $|F(q)|$ measurements versus magnetic field $B$ taken at $T = 130$ mK. The sample was rotated in $\Omega$ with respect to $\textbf{B}$ for each set. Note the small increase in $|F(q)|$ with increased $\Omega$. The data were fitted to $\ln(y) = a-bx$ to extract $\lambda$ and $c$ using equation~\ref{eq:FF_cc}, with the value of $\xi(T)\approx \xi_0(\Omega = 0^{\circ},10^{\circ}) = 20.3$ nm and $\xi_0(\Omega = 30^{\circ}) = 25.1$ nm at this temperature. } 
  
\end{figure}

When $\Omega = 0^{\circ}$, two VL domains with comparable signals are clearly present.  On rotating $\Omega$, we preferentially select one domain (illustrated in Fig.~\ref{fig:0p4T_0p1K}).  We might normally expect the overall $\langle I(q) \rangle$, summing over both domains, to be conserved for small $\Omega$ rotation, or even to decrease due to the $a$-$c$ anisotropy.  This is not the case here, as shown in Fig.~\ref{fig:FF_B_pub_log}, where there is a consistent increase in $\langle I(q) \rangle \propto |F(q)|^2$ as $\Omega$ increases.  This may be due to a decrease in the disorder of the VL. We have previously shown evidence in Figure~\ref{fig:ETA_average} to suggest that a rotation by $\Omega = 30^{\circ}$ is sufficient to `cancel-out' non-local and/or anisotropic effects. 

The orientation of the VL favored by $\Omega$ rotation is also of interest: the theory in the London approximation has been discussed by Campbell \emph{et al.}~\cite{Campbell88}. They predict that the the preferred VL orientation should give a pattern containing diffraction spots top and bottom; this is the opposite of what we observe, indicating higher-order contributions to the anisotropy. 

We now consider the variation with field of the form factor for the different $\Omega$ values as shown in Figure \ref{fig:FF_B_pub_log}. To allow for the presence of two VL domains occupying the sample volume, the average intensity from the two domains has been added to calculate the form factor. The straight lines fitted to the linear regions of these plots use a modified London model ~\cite{Tinkham96}, which has been shown experimentally to work well at low temperature for $B$ well below $B_{\rm c2}$~\cite{White11}:
\begin{equation}
|F(q)| = \frac{B}{1 + q^2 \lambda^2} e^{-c q^2 \xi^2}.
\label{eq:FF_cc}
\end{equation}
Here, $c$ is a constant in the Gaussian cut-off term that, along with the coherence length $\xi$, represents the effects of overlap of finite-width cores. Under the conditions of our measurements, $B >> B_{\textrm c1}$, the expression outside the exponential is essentially constant. 

\begin{figure}
    \includegraphics[width=0.48\textwidth]{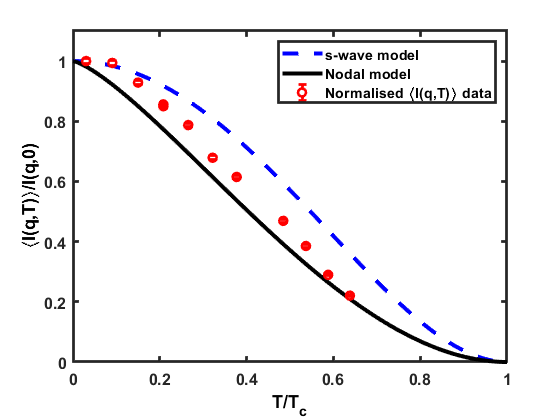}
  \caption{\label{fig:II_fitting} The temperature dependence of the integrated intensity of the dominant domain at $\Omega = 30^{\circ}$, with an applied field of 0.15 T. normalized to the extrapolated value at 0 K (this was calculated from the fits in Figure 9 and found to be very close to the maximum $I(q)$ value in the temperature dependent data sets).  The measured values are compared with models for \emph{s}-wave (dashed line) and nodal (solid line) superconductivity using the Prozorov \cite{Prozorov06} framework as described in the main text.}
\end{figure}

\begin{figure}
    \includegraphics[width=0.48\textwidth]{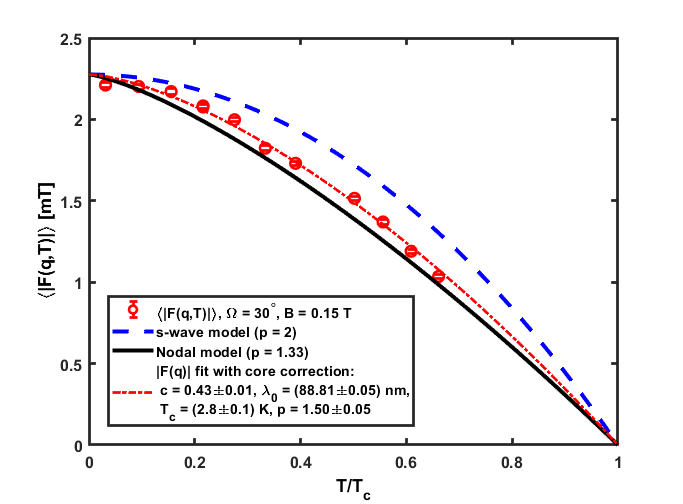}
  \caption{\label{fig:FF_fitting}The average form factor $\langle|F(q,T)|\rangle$ (red circles) is compared with ideal models for \emph{s}-wave (dashed line) and nodal (solid line) gap structures. The thin dashed line represents a fit to $\langle|F(q,T)|\rangle$ with variable $p$ in Eq.~4 inserted in Eq.~\ref{eq:FF_cc}. The fit parameters for this are given in the panel.}
\end{figure}

Because the measured signal from one domain was strongest at $\Omega = 30^{\circ}$, a detailed temperature dependence was measured in this condition at the low field of 0.15 T, which gives a strong intensity and will have reduced effects of vortex core overlap. Figure \ref{fig:II_fitting} shows the normalized integrated intensity versus temperature, along with theoretical lines that will be described later. From these data, the temperature-dependence of the average form factor, $\langle|F(q,T)|\rangle$, is obtained using equation~\ref{eq:Christen}. This is plotted in Figure \ref{fig:FF_fitting}.

The form factor can then be used to obtain the temperature-dependence of the penetration depth, $\lambda(T)$, using equation~\ref{eq:FF_cc}. This is very robust in the low temperature regime with a temperature-independent value of $\xi(T) \approx \xi_0$. This Brandt approach \cite{Brandt72, Brandt95, Forgan02} is justified by the largely linear behaviour of the field-dependent results in Figure~\ref{fig:FF_B_pub_log} and by the fact that we are operating in the low temperature regime. The value of $\xi_0$ was estimated from our heat capacity results as described earlier. The constant value of $c$ was determined by fitting the field dependent data in Figure~\ref{fig:FF_B_pub_log} with equation~\ref{eq:FF_cc}. The value for $T_c(B)$ was also obtained from the heat capacity measurements presented in Figure~\ref{fig:CT_with_B_c_2}. This was used in the fitting and all calculations using the $I(q)$ data. For 0.15 T applied field $T_c(B) = 2.8\pm0.1$ K.

Fits are available using BCS theory~\cite{Prozorov06, Lewis56, Mao95} to model $\lambda(T)$, but these rely on the assumption of a spin-singlet \emph{s}-wave gap structure. This material has already been established as a moderately heavy fermion system, and was thought to have two gaps \cite{Hong14}, so a BCS expression may not suffice. Instead we fit $\lambda(T)$ with a simple phenomenological expression which gives a good representation of its temperature-dependence, in order to discover what the fitting parameters indicate regarding the gap structure. We follow the approach developed in Refs.~\cite{Prozorov06, Lewis56, Mao95}, which is a simplification of the work by Izawa \emph{et al.} \cite{Izawa02, Maki02}. This uses an extension of the phenomenological Lewis two-fluid model for $\lambda(T)$ \cite{Lewis56, Tinkham96},
 
\begin{equation}
\lambda(T) = \frac{\lambda(0)}{\sqrt{1-t^4}},
\label{eq:lambda_BCS}
\end{equation}
where $t = T/T_c$. This was originally intended to represent a clean local BCS superconductor, but an exponent of 4 seems not to be appropriate even for that. Hence it was applied to more general situations by Prozorov \emph{et al.}~\cite{Prozorov06}, by introducing a variable exponent:

\begin{equation}
\lambda(T) = \frac{\lambda(0)}{\sqrt{1-t^p}}.
\label{eq:lambda_fit}
\end{equation}

Fits to the BCS theory results have shown that $p=2$ is a better representation of $s$-wave behaviour~\cite{Tinkham96, Lewis56}. $p = 4/3$ has similarly been shown to fit for the nodal $d$-wave gap structure \cite{Prozorov06}.

\begin{figure}
      \includegraphics[width=0.48\textwidth]{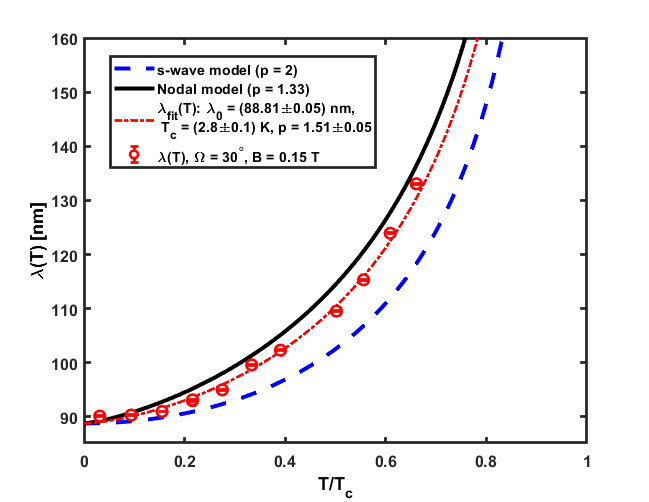}
  \caption{\label{fig:Lambda_fitting} $\lambda(T)$ versus temperature, calculated using Eq.~\ref{eq:FF_cc} rearranged, with $c = 0.65$ (see Figure~\ref{fig:FF_B_pub_log}).  The lines correspond to the ideal Prozorov models for \emph{s}-wave (dashed line) and nodal (solid line) superconductivity using Eq.~\ref{eq:lambda_fit}. A fit to the variable-$p$  Prozorov model is shown by the red short dashed line.}
\end{figure}

We then generate models for $\langle|F(q,T)|\rangle$ and $\langle I(q,T) \rangle$ for various values of $p$ and see how the empirical results compare. This approach helps classify the pairing symmetry of the gap function and potentially highlights any suppression of specific pairing mechanisms based on changes in $p$. The models created with this method are given in Figures~\ref{fig:II_fitting},~\ref{fig:FF_fitting} and~\ref{fig:Lambda_fitting} for $\langle I(q,T) \rangle$, $\langle|F(q,T)|\rangle$ and $\lambda(T)$, respectively.

In Figure~\ref{fig:II_fitting}, we see that the temperature-dependence of the integrated intensity lies between the nodeless and nodal models, although closer to the nodal value of $p$. In Figure~\ref{fig:FF_fitting} we have converted the intensity to form factor and also performed a fit with $p$ allowed to vary. We used the core correction value of $c = 0.43 \pm 0.01$ obtained with $\Omega = 30^\circ$. A good fit was obtained with $p = 1.50\pm0.05$, intermediate between nodeless and nodal values.

Alternatively, the results may be converted into penetration depth, using equation~2, and this is shown in Figure~\ref{fig:Lambda_fitting} , with  equation~4 as the fitting function. In this case also, the fitted value of $p=1.50\pm0.05$ suggests a fairly consistent tendency of the temperature-dependence of $\lambda$ towards the nodal model ($p=1.33$), rather than the s-wave model ($p = 2$).

\section{Discussion}

If we were seeing multiple gaps - two gaps as proposed previously \cite{Wang13, Hong14} - we might expect to see evidence of one or more of these gaps being suppressed at some field below $H_{c_2}$. Previous work identified a feature in the thermal conductivity that put a smaller gap being suppressed at $\approx$0.29 T. We do not see any sudden shift in the form factor signal around this field, nor do we see a sudden shift in the VL structure or anisotropy in the vicinity of this field in the form factor. However, this does correspond to the crossover of the opening angle, $\eta<60^{\circ} \rightarrow \eta>60^{\circ}$ in Figure~\ref{fig:ETA_average} for the low-angle results. It is possible this feature is very smooth, with a smooth transition over the suppression point of the smaller gap. Indeed, we do see that field dependent anisotropy is weak, but present, in this material. This is often evidence of multiple gaps due to the differing sensitivity of the gaps on each FS sheet~\cite{Kuhn16}, creating direction-dependent strength of the supercurrents in the crystal. 

In previous work TlNi$_2$Se$_2$ showed some evidence of potentially being a \emph{d}-wave superconductor \cite{Wang13}. Generally speaking \emph{d}-wave superconductivity can be identified in SANS studies by a change of the VL structure with field or angle~\cite{Bianchi08}. In this investigation we have seen no such rearrangement. This does not preclude the existence of \emph{d}-wave pairing entirely, but it is far less likely. Anisotropy is small in the VL with field and angle variation but has a consistent relationship with field variation and reflects a possible shift of flux lines attempting to align with the fourfold crystal structure. Previous photoemission and Raman spectroscopy measurements by Xu \emph{et al} demonstrate van Hove singularities (VHS) with fourfold symmetry about the \emph{Z} point in the FS~\cite{Xu15} (these VHS are held as the explanation for the observed heavy-fermion behaviour). We also see that by rotation of the crystal, the anisotropy can be minimised.

Given the empirical fits of $|F(q)|$, the most likely candidate for the gap structure is a nodal \emph{s}-wave gap due to the consistent fits of $\lambda(T)$ and $|F(q)|$ to $p<2$, as outlined by Prozorov \emph{et al.} \cite{Prozorov06, Gross86}. The conspicuous lack of structural changes in the VL is unusual for an unconventional superconductor and likely indicates we are not looking at a \emph{d}-wave pairing system~\cite{Bianchi08}. This combines to form a picture of a nodal or multigap with a small minimum gap, \emph{s}-wave system. This is somewhat concurrent with the conclusions of~\cite{Hong14}, which supported a nodeless, multigap system.

\section{Conclusion}

We can conclude that TlNi$_2$Se$_2$ is likely a nodal or small minimum gap, \emph{s}-wave superconductor, given the behaviour of the form factor supported by the applied empirical fits. Due to the observed weak anisotropy and lack of rearrangement of the VL we cannot attribute \emph{d}-wave behaviour. Although there is a  lack of features in the vicinity of the predicted suppression field, $H^* = 0.29$ T in field dependent $|F(q,B)|$ results, there is a cross-over of the VL opening angle at $\approx 0.26$ T for field directions close to the {\bf c}-axis. This would suggest a small anisotropy in the system concurrent with a multigap description of the pairing in TlNi$_2$Se$_2$.

Continued investigation of this material will clarify some of the unusual results given in this work. It would be prudent to continue SANS studies of the VL up to much larger angles of rotation of the field with respect to the crystal {\bf c}-axis, in order to probe for any structural changes in the VL and to see how the form factor signal continues to evolve with angle. In addition, it would be of interest to investigate how the VL anisotropy changes at larger values of applied magnetic field.

\section{Acknowledgements}

This work was supported by the U.K.~Engineering and Physics Sciences Research Council (EPSRC) funding under award No. 1521657 and grant No. EP/J016977/1. This work was is based on experiments performed at the Institute Laue Langevin (ILL). We are grateful for support from the National Basic Research program of China under Grant No.~2016YFA0300402, 2015CB921004 and the National Natural Science Foundation of China (No.~11374261), the Zhejiang Provincial Natural Science Foundation (No.~LY16A040012) and Fundamental Research Funds for the Central Universities of China.

\end{document}